\documentclass{elsarticle}

\usepackage{amsmath}
\usepackage{amssymb}
\usepackage{graphicx}
\usepackage{color}

\begin{document}

\title{Error correction in short time steps during the application of quantum gates}

\author{L. A. de Castro}
\ead{leonardo.castro@usp.br}

\author{R. d. J. Napolitano}

\address{S\~{a}o Carlos Institute of Physics, University of S\~{a}o Paulo \\ PO Box 369, 13560-970, S\~{a}o Carlos, SP, Brazil}

\begin{abstract}
We propose a modification of the standard quantum error-correction method to enable the correction
of errors that occur due to the interaction with a noisy environment during quantum gates
without modifying the codification used for memory qubits. Using a perturbation treatment of the
noise that allows us to separate it from the ideal evolution of the quantum gate, we demonstrate
that in certain cases it is necessary to divide the logical operation in short time steps intercalated
by correction procedures. A prescription of how these gates can be constructed is provided, as well
as a proof that, even for the cases when the division of the quantum gate in short time steps is not
necessary, this method may be advantageous for reducing the total duration of the computation.
\end{abstract}

\maketitle
\section{Introduction}

Since the beginning of quantum-computation theory,
it has been known that decoherence and other forms of external interference in the quantum bits (qubits) pose a difficulty in implementing real quantum computers~\cite{Unruh1994}.
Various methods have been devised to cope with these errors, including the quantum error-correction theory, which consists of encoding quantum states of $k$ logical qubits in $n>k$ physical qubits, thus generating a redundancy of information that can be used to identify and correct errors~\cite{ShorSteane}. Many of these codes have since been designed for different kinds of quantum channels.

It has been shown that quantum codes can be employed in quantum circuits as long as the gates are fault-tolerant~\cite{Shor1996}. The traditional fault-tolerant methods consider that the errors can be described as happening after the physical gates are applied~\cite{Steane1998}, which is a harmless assumption when dealing with a code capable of correcting any generic error up to a certain number of qubits. In this case, any kind of error that may emerge from the simultaneous interaction of the gate and the environment will still be correctable.

This assumption, nevertheless, does not remain true if we take into account quantum gates that have finite-time durations which are not negligibly small in comparison with the decoherence time~\cite{Gupta1999}. When we let both the errors and the gates occur simultaneously, unexpected new kinds of errors may affect the encoded qubits, which may render the correction procedure impractical if we are using a code that was designed to correct the noise of a specific quantum channel. Given that such channel-adapted codes can be more efficient than generic ones \cite{Fletcher2008}, a method of correcting errors that occur during the gate using the same specific code developed for the case of a memory qubit deserves attention.

In this article, we propose an alternative method of correcting errors caused by an external environment occurring during a logical operation that consists of repeatedly applying the   quantum error-correction procedure for memory qubits. Employing the analytic theoretical methods of standard quantum coding theory and of the dynamical evolution of a statistical quantum ensemble given by the von-Neumann equation, we lay down the conditions for the repeated error correction to be necessary. We also show how it can be applied to correct a universal set of quantum gates, exemplified in the case of the three-qubit phase-error-correcting code~\cite{Ekert1998}.

Special attention is given to the form such gates must be constructed. While fault-tolerant circuits employ transversal gates, which have been proven incapable of constructing a universal set for codes that correct a general one-qubit error~\cite{Eastin2009}, our method bypasses such restrictions by employing gates which cannot be factored in a tensor product of individual-qubit operations, being therefore non-transversal. The use of such gates does not cause additional errors with significant probability, but may require some special arrangement, as explained in the body of the article. 

This article is structured as follows: in Sec. II, we examine
the essential mathematics necessary to understand
the small-step method. In Sec. III, we explain how the
method works, and in Sec. IV we show how the gates can
be implemented within this framework. Sec. V is dedicated
to describe how the gates can be implemented by
an approximative method when multiple-qubit interactions
are prescribed. We conclude in Sec. VI, presenting
some further perspectives of work.
\section{Mathematical formulation}

In the description of a quantum computer prone to errors, we consider that the qubits are an object-system $S$ (to employ the same terminology as \cite{Everett}) that interacts with an environment $E$, usually modeled as bosonic baths.  Even though  we are interested only in the reduced density matrix of the qubits $\rho_S(t) = \mathrm{Tr}_E \left\{ \rho_{SE} (t) \right\}$, we must take into account the whole system and environment, described by the total density operator $\rho_{SE}(t)$, so we can describe a unitary evolution given by the von Neumann equation:

\begin{equation*}
\frac{\mathrm{d}}{\mathrm{d} t} \rho_{SE}(t) = - i \left[ H, \rho_{SE}(t) \right],
\end{equation*}

\noindent where $H$ is the Hamiltonian operating on both object-system and environment and we chose natural units, so that $\hbar =1$.

The solution of this equation is found to be 

\begin{equation}
\rho_{SE}(t) = U_{SE}(t) \rho_{SE}(0) U_{SE}^\dagger(t), \label{evolution}
\end{equation}

\noindent where the time-evolution operator $U_{SE}(t)$ is $e^{-iHt}$ if the Hamiltonian is time-independent. This Hamiltonian can always be split in three terms: two that act on the object-system and environment separately ($H_S$ and $H_E$) and one term of interaction that acts on both ($H_{\mathrm{int}}$). The latter must be weak enough for the errors to be rare -- otherwise, the recovery of the original state will be impossible. To make this assumption explicit, we write the interaction term multiplied by a small number $\lambda$:

\begin{equation*}
H = H_S + H_E + \lambda H_{\mathrm{int}}.
\end{equation*}

To treat the evolution due to the interaction with the environment as a perturbation with respect to the case of an error-free system, we expand the time-evolution operator $U_{SE} (t)$ in a power series of $\lambda$:

\begin{eqnarray*}
U_{SE}(t,\lambda) &=& U_{SE} (t,\lambda=0)
+ \lambda \left. \frac{\partial}{\partial \lambda} U_{SE} (t,\lambda) \right|_{\lambda=0} \\
&& + \frac{1}{2} \lambda^2  \left. \frac{\partial^2}{\partial \lambda^2} U_{SE} (t,\lambda) \right|_{\lambda=0} + \\
&& + \frac{1}{3!} \lambda^3 \left. \frac{\partial^3}{\partial \lambda^3} U_{SE} (t,\lambda) \right|_{\lambda=0} + \ldots,
\end{eqnarray*}

\noindent which can be re-written as

\begin{equation}
U_{SE}(t) = U_S(t) U_E(t) \left[ 1 + \sum_{n=1}^\infty \lambda^n C_n(t) \right] , \label{expansion}
\end{equation}

\noindent where the $C_n(t)$ are operators that can be found by:

\begin{equation}
C_n(t) = U_S^\dagger (t) U_E^\dagger (t) \left. \frac{1}{n!}  \frac{\partial^n}{\partial \lambda^n} U_{SE}(t) \right|_{\lambda=0}. \label{Cn}
\end{equation}
\noindent The operators $U_S(t)$ and $U_E(t)$ represent the evolution due to the object-system and the environment Hamiltonians alone, respectively. It is clear that, when $\lambda=0$, the Hamiltonian will be the sum of two commuting terms, $H_S$ and $H_E$. The complete time evolution operator, therefore, will be the product of $U_S(t)$ and $U_E(t)$, where the former is the evolution we would find if only the object system were contributing
\begin{equation*}
\frac{\mathrm{d}}{\mathrm{d}t} \left\{ U_S(t) \rho(0) U_S^\dagger \right\} =
-i \left[ H_S, U_S(t) \rho(0) U_S^\dagger(t) \right],
\end{equation*}
\noindent and, for the latter, if just the environment contributes:
\begin{equation*}
\frac{\mathrm{d}}{\mathrm{d}t} \left\{ U_E(t) \rho(0) U_E^\dagger (t) \right\} =
-i \left[ H_E, U_E(t) \rho(0) U_E^\dagger (t)\right] .
\end{equation*}

It is clear that the joint evolution $U_{SE}(t,\lambda=0)$ is simply the product of these two operators, which can be verified by taking the derivative:
\begin{eqnarray*}
\lefteqn{ \frac{\mathrm{d}}{\mathrm{d}t}
\left\{ U_S(t) U_E(t) \rho(0) U_E^\dagger (t) U_S^\dagger (t) \right\} = } \\
&& -i \left[ H_S + H_E, U_S(t) U_E(t) \rho(0) U_E^\dagger (t) U_S^\dagger (t) \right].
\end{eqnarray*}

Replacing the power expansion of the operator $U_{SE}(t)$ given in Eq.~(\ref{expansion}) in the final state of the system from Eq.~(\ref{evolution}), we find, up to the second order in $\lambda$:

\begin{eqnarray}
\rho_S(t) &=& U_S(t) U_E(t) \rho_S(0) U_S^\dagger (t) U_E^\dagger (t) \nonumber \\
&& + \lambda U_S(t) U_E(t) \rho_{SE} (0) C_1^\dagger (t) U_S^\dagger (t) U_E^\dagger (t) \nonumber \\
&& + \lambda U_S(t) U_E(t) C_1(t) \rho_{SE}(0) U_S^\dagger (t) U_E^\dagger (t) \nonumber \\
&& + \lambda^2 U_S(t) U_E(t) C_1(t)  \rho_{SE} (0) C_1^\dagger (t) U_S^\dagger (t) U_E^\dagger (t) \nonumber \\
&& + \lambda^2 U_S(t) U_E(t) \rho_{SE} (0) C_2^\dagger (t) U_S^\dagger (t) U_E^\dagger (t) \nonumber \\
&& + \lambda^2 U_S(t) U_E(t) C_2(t) \rho_{SE}(0) U_S^\dagger (t) U_E^\dagger (t) \nonumber \nonumber \\
&& + O(\lambda^3), \label{full_size}
\end{eqnarray}

\noindent If we impose the condition that $U_{SE}(t)$ is unitary in Eq. (\ref{expansion}), that is, $U_{SE}^\dagger (t) U_{SE}(t) = 1$, we must conclude that
\begin{equation*}
\left[ 1 + \sum_{n=1}^\infty \lambda^n C_n^\dagger(t) \right]
\left[ 1 + \sum_{n=1}^\infty \lambda^n C_n(t) \right] =1.
\end{equation*}
\noindent By equating the coefficients of each power of $\lambda$ on both sides of the equation, we find that
\begin{eqnarray*}
C_1^\dagger (t) &=& - C_1 (t), \\
C_2^\dagger (t) &=& -C_1^\dagger (t) C_1 (t) - C_2 (t).
\end{eqnarray*}

Therefore, taking the partial trace over the degrees of freedom of the environment in Eq. (\ref{full_size}), we have:

\begin{eqnarray}
\rho_S(t) &=& U_S(t) \rho_S(0) U_S^\dagger (t) \nonumber \\
&& + \lambda  U_S(t) \mathrm{Tr}_E \left\{ \left[ C_1(t) , \rho_{SE} (0) \right] \right\} U_S^\dagger (t) \nonumber \\
&& + \lambda^2  U_S(t) \mathrm{Tr}_E \left\{ \left[ \rho_{SE}(0) C_1(t), C_1(t) \right] \right\} U_S^\dagger (t) \nonumber \\
&& + \lambda^2  U_S(t) \mathrm{Tr}_E \left\{ \left[ C_2(t) , \rho_{SE}(0) \right] \right\} U_S^\dagger (t) \nonumber \\
&& + O(\lambda^3), \label{pre_final}
\end{eqnarray}

\noindent where $\rho_S(t)$ is the reduced density matrix of the object system.

Applying Snider's identity~\cite{Wilcox} to calculate the derivative of the time-evolution operator, we find:

\begin{eqnarray*}
\frac{\partial}{\partial \lambda} U_{SE}(t,\lambda) &=& -i U_{SE}(t,\lambda) \\
&& \times \left[ \int_0^t \mathrm{d}t^\prime\; U_{SE}^\dagger (t^\prime,\lambda) H_{\mathrm{int}} U_{SE} (t^\prime,\lambda) \right].
\end{eqnarray*}

\noindent The exact origin of this formula is explained in Appendix A. Replacing  it at Eq. (\ref{Cn}) allows us to calculate both $C_1(t)$ and $C_2(t)$:

\begin{eqnarray*}
C_1(t) &=& U_S^\dagger (t) U_E^\dagger (t) \left. \frac{\partial}{\partial \lambda} U_{SE}(t,\lambda) \right|_{\lambda=0} \\
&=& -i \int_0^t \mathrm{d}t^\prime \; U_S^\dagger (t^\prime) H_I(t^\prime) U_S (t^\prime) \\
&=& -i \int_0^t \mathrm{d}t^\prime \tilde H_I(t^\prime)
\end{eqnarray*}
\noindent and
\begin{eqnarray*}
C_2(t) &=&  \frac{1}{2} U_S^\dagger (t) U_E^\dagger (t) \left. \frac{\partial^2}{\partial \lambda^2} U_{SE}(t,\lambda) \right|_{\lambda=0} \\
&=& -i \frac{1}{2} U_S^\dagger (t) U_E^\dagger (t) \frac{\partial}{\partial \lambda} \left[ U_{SE}(t,\lambda) \right. \\
&& \left. \times \int_0^t \mathrm{d}t^\prime\; U_{SE}^\dagger (t^\prime,\lambda) H_{\mathrm{int}} U_{SE} (t^\prime,\lambda) \right]_{\lambda=0} \\
&=& \frac{1}{2} C_1(t) C_1(t) - \frac{1}{2} \int_0^t \mathrm{d}t^\prime \; \int_0^{t^\prime} \mathrm{d}t^{\prime\prime} \\
&& \left[ U_S^\dagger (t^\prime) H_I (t^\prime) U_S(t^\prime), U_S^\dagger (t^{\prime\prime}) H_I (t^{\prime\prime}) U_S (t^{\prime\prime}) \right] \\
&=& -\frac{1}{2} \int_0^t \mathrm{d}t^\prime  \left\{ \int_0^t \mathrm{d}t^{\prime\prime} \tilde H_I(t^\prime) \tilde H_I(t^{\prime\prime})  \right. \\
&& \;\;\;\;\; \left. + \int_0^{t^\prime} \mathrm{d}t^{\prime\prime} \left[ \tilde H_I(t^\prime), \tilde H_I(t^{\prime\prime}) \right] \right\}
\end{eqnarray*}

\noindent where $H_I(t)\equiv U_E^\dagger (t) H_{\mathrm{int}} U_E(t)$ is the interaction-picture Hamiltonian for the memory, when there is no gate acting on the qubits; and  $\tilde H_I(t) \equiv U_S^\dagger(t) H_I(t) U_S(t)$ is the interaction-picture Hamiltonian for the case when the gate is being applied.

Therefore, the final state as written in Eq. (\ref{pre_final}) can be re-expressed as:

\begin{eqnarray*}
\rho_S(t) &=& U_S(t) \int_0^t \mathrm{d}t^\prime \mathrm{Tr}_E  \left\{ -i \lambda \left[ \tilde H_I(t^\prime) , \rho_{SE} (0) \right]  \right.  \\
&& \qquad - \frac{1}{2} \lambda^2  \int_0^t \mathrm{d}t^{\prime\prime}\left\{ \tilde H_I(t^\prime) \tilde H_I(t^{\prime\prime}),  \rho_{SE} (0) \right\} \\
&& \qquad - \frac{1}{2} \lambda^2  \int_0^{t^\prime} \mathrm{d}t^{\prime\prime} \;\left[ \left[ \tilde H_I (t^\prime),  \tilde H_I (t^{\prime\prime}) \right] , \rho_{SE}(0) \right]  \\
&& \left. \qquad + \lambda^2 \int_0^t \mathrm{d}t^{\prime\prime} \tilde H_I(t^\prime) \rho_{SE} (0) \tilde H_I(t^{\prime\prime})\right\} U_S^\dagger (t) \\
&& +U_S(t) \rho_S(0) U_S^\dagger (t) + O(\lambda^3).
\end{eqnarray*}

\noindent The first-order term is usually neglected~\cite{BreuerPetruccione}, which can be achieved by the imposition that $ \mathrm{Tr}_E \left\{ H_I(t^\prime) \rho_E (0) \right\} = 0$. This condition is especially valid when the environment is a bosonic bath, in which case the interaction Hamiltonian usually contains terms with a single annihilation ($a_k$) or creation ($a_k^\dagger$) operator:

\begin{equation}
H_{\mathrm{int}} = \sum_{k} S_{k} a_{k} + \text{h.c.}, \label{formamiltoniana}
\end{equation}

\noindent where the $\{ S_{k} \}$ represent operators that act only on the system. As the trace of a single creation or annihilation operator always vanishes:

\begin{equation*}
\mathrm{Tr}_E \{ a_k \} = \mathrm{Tr}_E \{ a_k^\dagger \} = 0,
\end{equation*}

\noindent the expectation value of the interaction Hamiltonian will be zero if the initial state of the bath is a Boltzmann distribution such as $e^{-\hbar\beta\sum_{k}\omega_{k}a_{k}^{\dagger}a_{k}}$. This is not necessarily true for every state of the environment, however, so the condition $\mathrm{Tr}_{E}\left\{ H_{I}\left(t^{\prime}\right)\rho_{E}\left(0\right)\right\} =0$ must always be independently assumed in order to eliminate the first-order term.

A similar reasoning is used for every term containing an odd number of Hamiltonians, leaving only the even powers of $\lambda$ in the expansion.

Considering the state of the object-system initially separable from the environment -- before $t=0$ no errors were occurring, which means that $S$ and $E$ cannot be entangled -- we find the following equation for the final state:

\begin{eqnarray}
\rho_S(t) &=& \frac{1}{2} \lambda^2 t^2 U_S(t) \mathrm{Tr}_E \left\{ \mathcal{E}_S (t) \rho_{SE} (0) \right\} U_S^\dagger(t) \nonumber \\
&& + U_S(t) \rho_S(0) U_S^\dagger (t) + O(\lambda^4), \label{evolution2}
\end{eqnarray}

\noindent where the error superoperator $\mathcal{E}_S (t)$ acts on the density matrix according to:

\begin{eqnarray*}
\mathcal{E}_S (t) \rho &\equiv& \int_0^1 du \left\{ \int_0^1 dv  \left[ \tilde H_I(ut) ,  \left[ \rho,  \tilde H_I(vt) \right] \right] \right. \\
&& \left. - \int_0^{u} dv \left[ \left[ \tilde H_I (ut) ,  \tilde  H_I (vt) \right] ,  \rho \right] \right\}.
\end{eqnarray*}

The second term on the right-hand side of Eq.~(\ref{evolution2}) represents how the system evolves when it is subject only to the gate Hamiltonian $H_S$. Therefore, the superoperator $\mathcal{E}_S (t)$ represents the errors that occur, with a probability proportional to $\lambda^2 t^2$, while the gate is being applied. However, this superoperator depends on the type of the gate. An arbitrary quantum code should not be designed to fix every kind of error, just those that occur with a probability proportional to $\lambda^2$ when the qubit is in the memory.

As our objective is to avoid the impractical case where a different code is used for each gate applied, the errors that occur during the gate must be of the same type as those that occur in the memory. The latter will be denoted by the superoperator $\mathcal{E}_0(t)$, which is the same as $\mathcal{E}_S(t)$, but with the gate Hamiltonian $H_S$ set to zero:

\begin{eqnarray*}
\mathcal{E}_0 (t) \rho &\equiv& \left. \mathcal{E}_S (t) \rho \right|_{H_S=0} \\
&=& \int_0^1 du \left\{ \int_0^{1} dv  \left[  H_I(u t) ,  \left[ \rho,  H_I(v t) \right] \right] \right. \\
&& \left. - \int_0^{u} dv \left[ \left[ H_I (u t) ,  H_I (v t) \right] ,  \rho \right] \right\}.
\end{eqnarray*}

\section{The short time step method}

A sufficient condition for the errors during the quantum gate to be correctable is the commutativity of the gate operator $U_S(t)$ and the interaction-picture Hamiltonian $H_I(t)$. In this case, the interaction Hamiltonian is identical to the case without gate ($\tilde H_I (t) =H_I(t)$), so that $\mathcal{E}_S(t)=\mathcal{E}_0(t)$ and the errors that occur are of the same type that happen in the memory. This fact is also evident if we expand the time-evolution operator in a part that represents the gate and another that represents the errors:

\begin{equation*}
U_{SE}(t) = e^{-i(H_S+H_E+H_{\mathrm{int}})t} = U_S(t) e^{-i(H_E+H_{\mathrm{int}})t},
\end{equation*}

\noindent which is true if $[H_S,H_{\mathrm{int}}]=0$. In this case, the evolution of the qubits due to the gate is separable from the evolution due to the noise. Therefore, there is no need to modify the standard error-correction procedure.

However, when those two terms do not commute, the kinds of errors generated by $\mathcal{E}_S(t)$ may not be correctable by the code. To solve this problem without having to resort to more resource-consuming codes, we propose the application of the quantum-gate operation of total duration $t_g$ in $N$ short time steps $\Delta t= N^{-1} t_g$. After each of these steps, the standard error-correction procedure is applied, until the system reaches the desired final state. The ordinary error-correction procedure works in this method because, for small time intervals, the Hamiltonians $H_S$ and $H_I(t)$ commute approximately, resulting in an interaction-picture Hamiltonian identical up to the zeroth order in $N$ to the case when no gate is being applied:

\begin{equation*}
\tilde H_I (\Delta  t) = H_I(\Delta t) + O(N^{-1}).
\end{equation*}

As each term of the error superoperator contains two expressions of the kind $\tilde H_I (\Delta t)$, the final result for the expansion of $\mathcal{E}_S(\Delta t)$ is:

\begin{eqnarray*}
\mathcal{E}_S (\Delta t) \rho &=& \int_0^1 du  \left\{ \int_0^1 dv  \left[  H_I\left(u \Delta t \right) ,  \left[ \rho,   H_I\left( v \Delta t \right) \right] \right] \right. \\
&& \qquad \left. - \int_0^u dv \left[ \left[  H_I \left(u \Delta t \right) ,  H_I \left( v \Delta t \right) \right] ,  \rho \right] \right\} \\
&& + O(N^{-1}) \\
&=& \mathcal{E}_0 (\Delta t) \rho + O(N^{-1}).
\end{eqnarray*}

\noindent which leads to the following state of the object-system after the short time interval, according to Eq.~(\ref{evolution2}):

\begin{eqnarray}
\rho_S((n+1)\Delta t) &=&  \frac{\lambda^2 t^2}{2 N^2}  U_S(\Delta t) \mathrm{Tr}_E \left\{ \mathcal{E}_0 (\Delta t) \rho_{SE} (n\Delta t) \right\} U_S^\dagger(\Delta t) \nonumber \\
&& +  U_S(\Delta t) \rho_S(n \Delta t) U_S^\dagger (\Delta t) + O(\lambda^2 N^{-3}), \label{final_res}
\end{eqnarray}

\noindent where $n$ is a positive integer less than or equal to $N$. An error-correction proceeding is applied at each instant $t=n\Delta t$, resulting in a state proportional to $U_{S}\left(\Delta t\right)\rho_{S}\left(n\Delta t\right)U_{S}^{\dagger}\left(\Delta t\right)$  plus corrections of order $O\left(\lambda^{2}N^{-3}\right)$. After $N$ of these steps, we reach the desired reduced final state $U_{S}\left(N\Delta t\right)\rho_{S}\left(0\right)U_{S}^{\dagger}\left(N\Delta t\right)$ plus corrections of order $O\left(\lambda^{2}N^{-2}\right)$. That such corrections can be neglected is argued in the paragraphs that follow.

Notice that Eq.~(\ref{evolution2}) is only valid when the initial density matrix is separable in object-system and environment components. This is the case, we are assuming, in the beginning of the computation, right after the initial state has been prepared, at $t=0$. As time passes, in $t>0$, the equation in general does not hold anymore, because the process of error entangles the qubits with the environment. It only holds again at the instants $t=n\Delta t$, the instants just after a successful measurement and error correction procedure. These are capable of separating system and environment, thus rendering the state $\rho_{SE}\left(n\Delta t\right)$ as workable in the equation as $\rho_{SE}\left(0\right)$.

Choosing the number of steps $N$ as approximately $\lambda^{-1/2}$, the probability of finding an uncorrectable error is of the order of $\lambda^2 N^{-3} \sim \lambda^{7/2}$, negligible in comparison with the probability of finding a correctable error, which is represented by the first term in the right-hand side of Eq. (\ref{final_res}), which is of the order of $\lambda^2 N^{-2} \sim \lambda^3$. Therefore, the correction method designed for memory qubits can be applied during the gate with low probability of failure.

(The argument above rests on the assumption that $\mathcal{E}_0(\Delta t)$ is predominantly independent of $\Delta t$, so that no error terms of greater order than $\lambda^2 N^{-2}$ is introduced. Why this is generally justified can be seen in Appendix B.)

The repeated application of the error correction does not increase the probability of an uncorrectable error, originally of the order of $\lambda^3$: if the  gates are split in $N$ parts, each with a probability $\lambda^2 N^{-3}$ of suffering an uncorrectable error, the final probability of error is of the order $N \left( \lambda^2 N^{-3} \right)  \sim \lambda^3$, one order of magnitude smaller than correctable ones.

The probability of having to apply a correction procedure will be of the order of $\lambda^{2}$ if the gate is not divided in smaller steps. When we apply the gate in $N$ small time steps, the probability of having to apply a correction during its entire duration is of the order of $N\left(\lambda N^{-1}\right)^2 \sim\lambda^{5/2}$. Therefore, the number of times a correction procedure has to be applied is actually smaller than in the standard method. Whether our method causes a significant overhead in the processing, therefore, depends on the duration of the syndrome measurement and classical processing of its result, which must necessarily be applied $N$ times per gate.

The only difficulty we have not dealt with so far are the errors that may occur while the correction gate itself is being applied. However, one must note that the probability of a complete gate requiring a correction procedure (estimated above as of the order of $\lambda^2 N^{-1}$), and the correction procedure itself suffering some sort of error ($\lambda^2$, as in the case of an ordinary unprotected gate) yields a result of the order of $\lambda^4 N^{-1} \sim \lambda^{9/2}$. This is much less than the probability of an uncorrectable error (given above in $\lambda^3$), and, therefore, safely negligible.

It is important to note that even in the case when the partition of the gate in small steps may not be necessary (as when the logical gates all commute with the interaction Hamiltonian or when the unexpected errors are fortuitously correctable by the code), this method can still be useful to reduce the probability of an error happening and, therefore, the overall duration of the computation. Indeed, as demonstrated above, a quantum-gate operation with a probability of approximately $\lambda^2$ of having to be corrected, with this method is only required to be corrected with a probability of the order of $\lambda^{5/2}$. As the necessity of error corrections is smaller, the computation may be performed faster,  and the errors that occur while the correction gate is being applied can be safely ignored.

Additionally, we may note that this method does not require the excessive partition of the gate. If the error is of the order of $\lambda^2 \sim 10^{-4}$, then $\lambda \sim 10^{-2}$, and our prescription requires only $N \sim 10$ 
time steps. However, it is possible to increase the number of time steps to further reduce the probability of any error occurring. The decrease in the probability of error is only limited by experimental capability of partitioning the gates.

The caveat, however, is that our proposed method requires a gate $U_S(t)$ that keeps the state of the qubits inside the quantum code for every instant $t$, not only for the final instant $t_g$. Otherwise, the syndrome measurement will detect errors when none have happened. However, the construction of logical gates that respect this restriction does not pose an unsurmountable obstacle if we employ non-transversal gates explained in the next section.
\section{Gates}

Logical gates are intended to reproduce,  on an encoded logical state, the effects of some physical gate $G$ on a physical qubit state $\left| \psi \right\rangle$. Identifying the logical gates and states by the subscript $L$, if the physical gate acts according to
$G \left| \psi \right\rangle = \left| \psi^\prime \right\rangle$,
then the equivalent logical gate must act according to
$G_L \left| \psi \right\rangle_L = \left| \psi^\prime \right\rangle_L$.

All $n$-qubit quantum gates can be reproduced by some combination of CNOTs and single-qubit rotations~\cite{NielsenChuang}. The correct construction of these two gates, therefore, is sufficient to render any computation protectable by this method. Any rotation gate can be characterized by three angles $\theta$, $\phi$, and $\varphi$:

\begin{eqnarray*}
U_{\text{rot}} (\theta,\phi,\varphi) &=& \cos\varphi -i \sin\varphi \\
 && \times \left[ \sigma_z \cos\theta + \sigma_x \sin\theta \cos\phi + \sigma_y \sin\theta \sin\phi \right].
\end{eqnarray*}

\noindent The CNOT gate, on the other hand, has always the same structure:

\begin{equation*}
U_{\text{CNOT}} = \left( \frac{1+\sigma_{1,z}}{2} \right) + \left( \frac{1-\sigma_{1,z}}{2} \right) \sigma_{2,x},
\end{equation*}

\noindent where the index 1 represents the control qubit, while 2 represents the target.

The Hamiltonians used to generate these two gates are, for the rotation:

\begin{equation}
H_{\text{rot}} (\theta,\phi) = \omega_0 \left( \sigma_z \cos\theta + \sigma_x \sin\theta \cos\phi - i \sigma_x \sigma_z \sin\theta \sin\phi \right), \label{rot}
\end{equation}

\noindent which must be applied for a period $t_g= \omega_0^{-1} \varphi$, and, for the CNOT,

\begin{equation}
H_{\text{CNOT}} = \omega_0 \left( \frac{1-\sigma_{1,z}}{2} \right) \left(  \frac{1- \sigma_{2,x} }{2} \right), \label{CNOT}
\end{equation}

\noindent where the end time must be chosen so that $t_g=\omega_0^{-1} \pi$.

To create the equivalent logical gates, we must replace the physical bit flips $\sigma_x$ and phase flips $\sigma_z$ by their logical equivalents in Eqs. (\ref{rot}) and (\ref{CNOT}). In this case, the corresponding time evolution of the logical gate will retain the form of the physical gates as long as the logical Pauli matrices are Hermitian and still satisfy the identities

\begin{eqnarray}
\left( \sigma_{L,z} \right)^2 &=& \left( \sigma_{L,x} \right)^2 = 1, \label{Pauli1} \\
\sigma_{L,z} \sigma_{L,x} &=& -\sigma_{L,x} \sigma_{L,z}. \label{Pauli2}
\end{eqnarray}

\noindent The time-evolution operators for instants of time before the gate operation is complete will be combinations of the logical $\sigma_{L,x}$, $\sigma_{L,z}$ and  identity operators. As $\sigma_{L,x}$ and $\sigma_{L,z}$ are constructed with the intention of keeping the states inside the code subspace, both the rotation and the CNOT are guaranteed to keep the qubits inside the code for any instant of time.

Therefore, we just need to know how to define the logical Pauli matrices that satisfy Eqs.~(\ref{Pauli1}) and~(\ref{Pauli2}). We illustrate the construction of them for the triple-repetition phase-error-correcting code, defined as:

\begin{equation}
\left| m \right\rangle_L = \bigotimes_{k=1}^3 \left( \frac{1}{\sqrt{2}} \sum_{n=0}^1 (-1)^{mn} \left|n\right\rangle \right). \label{code}
\end{equation}

For this code, it can be easily seen that $\sigma_{1,x}$ fulfills the role of logical phase flip, and $\sigma_{1,z} \sigma_{2,z} \sigma_{3,z}$, the role of bit flip.
As these operators belong to the Pauli group, it is clear that they satisfy Eqs.~(\ref{Pauli1}) and~(\ref{Pauli2}), as required for the method to be applicable.
In this case, the partition of the gate in short time steps is also required, because the logical phase-flip gate $\sigma_{L,z}$ does not commute with the interaction-picture Hamiltonian for the phase error channel, which contains the Pauli matrix $\sigma_z$.

A final concern that must be addressed is that this method of constructing gates precludes the correction of errors due to an incorrect duration of the gate, besides requiring the use of Hamiltonians that may involve the interaction of three qubits. These problems can be circumvented if the gate itself is also simulated through the short-time application of simpler Hamiltonians, as explained in next section.
\section{Simulating the gates}

It can be seen that some of the gates needed by the method described in previous section include three-body interactions, which require non-physical Hamiltonians. To counter this problem, we can divide each of the multiple-body interactions in a series of two-body gates, which is valid for small time steps:

\begin{equation*}
e^{i \epsilon \sigma_{1,j_1} \sigma_{2,j_2} \ldots \sigma_{n,j_n} } = \prod_k e^{\alpha_k \sigma_{m_k, \ell_k} \sigma_{n_k, l_k} } + O(\epsilon^2).
\end{equation*}

\noindent The method used here to simulate the multiple-body Hamiltonians, inspired by those already developed for long periods~\cite{Tseng}, will take advantage of the fact that our gates are applied to small time-steps, so that the approximation given above is reasonable.

This scheme will be illustrated for the case of the three-qubit code from Eq.~(\ref{code}), in which case the $\sigma_{L,x}$ gate requires the simulation of a three-qubit Hamiltonian, while the CNOT requires a four-qubit one. In the first case, the time-evolution required is of the form $e^{-i \sigma_{1,z} \sigma_{2,z} \sigma_{3,z} \epsilon }$, which must be split into a product of exponentials that contain only two Pauli matrices. For this purpose, we will use the operators $e^{-i\sigma_{1,z} \sigma_{2,x}\epsilon }$ and $e^{-i\sigma_{2,y} \sigma_{3,z}\epsilon }$, chosen because the product of their arguments retrieves the Hamiltonian we are trying to simulate:

\begin{equation*}
\left( \sigma_{1,z} \sigma_{2,x} \right) \left( \sigma_{2,y} \sigma_{3,z} \right) = i \sigma_{1,z} \sigma_{2,z} \sigma_{3,z},
\end{equation*}

\noindent so that it is expected that a sequence of such operators will be capable of effectively simulating the unphysical gate we wish to apply.

\begin{figure}[thb]
\includegraphics[width=\columnwidth]{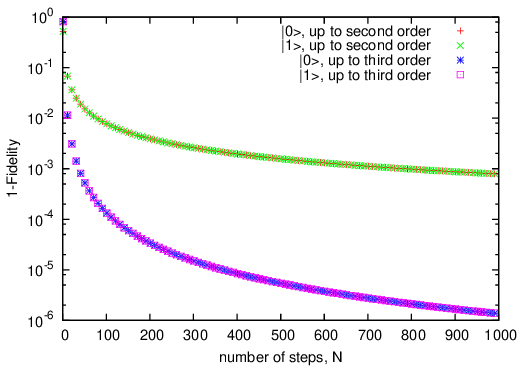}
\caption{\label{SigmaX}  Fidelity of the logical gate $\sigma_x$ for the three-qubit code, after split in physical two-body Hamiltonians. The two upper curves represent the two less efficient second-order approximations, while the two lower represent more precise third-order results. Both approximations were tested for two initial states, which correspond to the computational basis of the code subspace. The y-axis is in logarithmic scale to better represent the fast increase of the fidelity.}
\end{figure}

As these two matrices do not commute, we can expect that application in sequence of $e^{-i\sigma_{1,z} \sigma_{2,x}\epsilon } e^{-i\sigma_{2,y} \sigma_{3,z}\epsilon }$ and the same couple of operators with the sign of $\epsilon$ inverted will not yield a result proportional to the identity. Indeed, up to the second order in $\epsilon$, what we actually have is:

\begin{equation}
e^{-i\sigma_{1,z} \sigma_{2,x}\epsilon } e^{-i\sigma_{2,y} \sigma_{3,z}\epsilon } e^{i\sigma_{1,z} \sigma_{2,x}\epsilon } e^{i\sigma_{2,y} \sigma_{3,z}\epsilon } = 
1 - 2i \epsilon^2  \sigma_{1,z} \sigma_{2,z} \sigma_{3,z} + O (\epsilon^3), \label{expansionX2}
\end{equation}

\noindent which is equivalent to an approximation of the three-qubit Hamiltonian up to the second order in $\epsilon$. A numerical comparison of this simulation with the ideal non-physical gate is presented in Fig.~\ref{SigmaX}, where it can be seen that the order of the magnitude of the error will be $10^{-3}$ if the number of steps $N$ approaches $10^{3}$.

To improve on the quality of the simulation, we can expand Eq.~(\ref{expansionX2}) up to the third order in $\epsilon$, finding:

\begin{eqnarray*}
e^{-i\sigma_{1,z} \sigma_{2,x}\epsilon } e^{-i\sigma_{2,y} \sigma_{3,z}\epsilon } e^{i\sigma_{1,z} \sigma_{2,x}\epsilon } e^{i\sigma_{2,y} \sigma_{3,z}\epsilon } &=& 1 - 2i \epsilon^2  \sigma_{1,z} \sigma_{2,z} \sigma_{3,z} \\
&& +2i\epsilon^{3}\left(\sigma_{2,y}\sigma_{3,z}-\sigma_{1,z}\sigma_{2,x}\right)  + O (\epsilon^4).
\end{eqnarray*}

\noindent The additional term on the right-hand side can be turned into two physical exponentials (containing only two Pauli matrices) and put together with the other two-qubit terms, so that the expansion up the third order becomes:

\begin{eqnarray}
 e^{- 2i \epsilon^2  \sigma_{1,z} \sigma_{2,z} \sigma_{3,z}} &=& e^{-i\sigma_{1,z} \sigma_{2,x}\epsilon } e^{-i\sigma_{2,y} \sigma_{3,z}\epsilon }
e^{i\sigma_{1,z} \sigma_{2,x}\epsilon } e^{i\sigma_{2,y} \sigma_{3,z}\epsilon } 
e^{-2i\epsilon^{3}\sigma_{2,y}\sigma_{3,z}} e^{2i\epsilon^3\sigma_{1,z}\sigma_{2,x}} \nonumber \\
&& + O (\epsilon^4).  \label{SigmaX3}
\end{eqnarray}

The precision of this new simulation is also portrayed in Fig.~\ref{SigmaX}, showing that the threshold of $10^{-4}$ probability of error is achieved with just $N \sim 10^{2}$ steps.

The logic gate CNOT requires a four-qubit operator of the form $e^{-i \sigma_{1,z} \sigma_{2,z} \sigma_{3,z} \sigma_{4,x} \epsilon}$, the first expansion of which can be simply obtained by adapting Eq.~(\ref{SigmaX3}):

\begin{eqnarray}
e^{- 2i \epsilon^2  \sigma_{1,z} \sigma_{2,z} \sigma_{3,z} \sigma_{4,x}} &=& e^{-i\sigma_{1,z} \sigma_{2,z} \sigma_{3,x}\epsilon } e^{-i\sigma_{3,y} \sigma_{4,x}\epsilon } \nonumber \\
&& \quad \times e^{i\sigma_{1,z} \sigma_{2,z} \sigma_{3,x}\epsilon } e^{i\sigma_{3,y} \sigma_{4,x}\epsilon } \nonumber \\
&& \quad \times e^{-2i\epsilon^{3}\sigma_{3,y}\sigma_{3,x}} e^{2i\epsilon^3\sigma_{1,z} \sigma_{2,z}\sigma_{3,x}} \nonumber \\
&& + O (\epsilon^4). \label{CNOT4}
\end{eqnarray}

The problem now is to simulate the matrices of the type $e^{-i\sigma_{1,z} \sigma_{2,z} \sigma_{3,x}\epsilon }$, which are also non-physical. In order to keep their errors within the limits of this expansion, which correspond to the fourth order in $\epsilon^4$, an expansion up to the seventh order is required:

\begin{eqnarray*}
\lefteqn{ e^{i\left(2\epsilon^{2}-8\epsilon^{4}/3-56\epsilon^{6}/45\right)\sigma_{1,z}\sigma_{2,z}\sigma_{3,x}} = e^{i\epsilon\sigma_{1,z}\sigma_{2,x}}e^{-i\epsilon\sigma_{2,y}\sigma_{3,x}} } \nonumber \\
&& \times e^{-i\epsilon\sigma_{1,z}\sigma_{2,x}}e^{i\epsilon\sigma_{2,y}\sigma_{3,x}}e^{i\left(6\epsilon^{5}+16\epsilon^{7}/5\right)\sigma_{1,z}\sigma_{2,x}} \nonumber\\
&& \times e^{-i\left(2\epsilon^{5}-56\epsilon^{7}/5\right)\sigma_{2,y}\sigma_{3,x}}e^{-2i\epsilon^{3}\sigma_{1,z}\sigma_{2,x}} \nonumber \\
&& \times e^{-2i\epsilon^{3}\sigma_{2,y}\sigma_{3,x}} + O(\epsilon^8). \label{SigmaX7}
\end{eqnarray*}

Using the results of Eq.~(\ref{SigmaX7}) in conjunction with those of Eq.~(\ref{CNOT4}), we obtain a simulation of the CNOT that reaches the threshold of $10^{-4}$ for the probability of error when the number of steps is of the order of $10^2$, as illustrated in Fig.~\ref{CNOT}.

\begin{figure}[thb]
\includegraphics[width=\columnwidth]{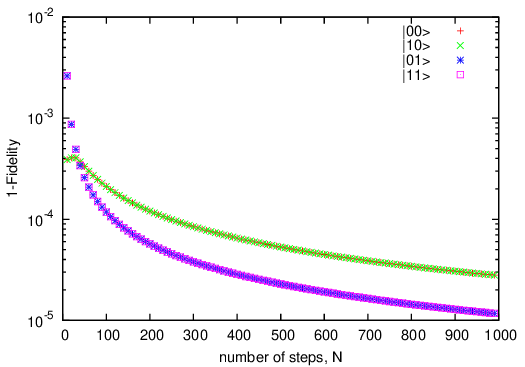}
\caption{ \label{CNOT} Fidelity of the logical gate $\text{CNOT}$ for the three-qubit code, after split in physical two-body Hamiltonians. The different curves represent  four different initial conditions, which are the computational basis for two encoded qubits. The difference between the two curves is due to the fact that, depending on the value of the control qubit (second qubit), the operation on the first qubit is distinct. Both curves have the same order of magnitude, nevertheless, as can be seen from the logarithm scale.}
\end{figure}

Should the experimental system require, the precision of the gates can be improved even more by the expansion of the simulation to even higher orders of magnitude. Likewise, it is not straightforward to repeat this procedure for a higher number of qubits. In so doing, we are not only able to simulate the multiple-body gate with two-body interactions, but we employ a series of operators that do not preserve the code subspace, rendering errors in the duration of the gate again detectable.

An additional aspect that can be considered is that this method of simulation need not be necessarily applied only in this context. When quantum simulations may be necessary, this method could be applied independently of the protection scheme to reproduce the effects of non-physical Hamiltonians.

\section{Conclusions}

We have demonstrated that applying the quantum gate in small steps intercalated by correction procedures can protect against external errors that happen during the quantum gate, even for a code designed for a specific quantum channel. This is achieved by decreasing the probability  of errors when these are incompatible with the code under normal circumstances. The only requirement of this method is the possibility of defining a set of logical Pauli matrices for the code that satisfy Eqs.~(\ref{Pauli1}) and~(\ref{Pauli2}). We have shown this to be possible for the simple three-qubit phase-error-correcting code, but it is expected that it can be adapted to any quantum code, as Eqs.~(\ref{Pauli1}) and~(\ref{Pauli2}) are derived from the operational definitions of the Pauli matrices.

Moreover, in this process we have demonstrated how the small-step method can be used to reduce the probability of even correctable errors, thus increasing the speed of the computation; and have presented a simulation method for the non-physical Hamiltonians that can be used independently of the error correction.

The main phenomenon of protection of quantum codes by repeated measurement, however, is not entirely unexpected: a syndrome measurement performed in close intervals restricts the probability of an error happening~\cite{Erez2004}. Other studies of continuous error correction \cite{Zurek} (and some related work from our group \cite{Brasil}) suggest that the limit of measurements performed continuously while the errors occur may protect even more effectively the qubits.

\section*{Acknowledgements}
L. A. de Castro acknowledges support from Funda\c c\~ ao de Amparo \` a Pesquisa do Estado de S\~ ao Paulo (FAPESP), Brazil, project number 2009/12460-0, and Coordena\c c\~ ao de Aperfei\c coamento de Pessoal de Ensino Superior (CAPES), Brazil. R. d. J. Napolitano acknowledges support from Conselho Nacional de Desenvolvimento Cient\' ifico e Tecnol\' ogico (CNPq), Brazil.
\section*{Appendix A: Derivation of Snider's formula}

Here we will see how to obtain the formula of the partial derivative for any unitary evolution operator $U(t)$.

We begin from the Schr\"{o}dinger equation. We will assume that the Hamiltonian it is associated with may be time-dependent:
\begin{equation*}
\frac{\mathrm{d}}{\mathrm{d}t} U(t) = -i H(t) U(t).
\end{equation*}

Next, we take the derivative of both sides of the equation with respect to some parameter $\lambda$:
\begin{equation*}
\frac{\partial}{\partial \lambda} \frac{\mathrm{d}}{\mathrm{d}t} U(t) =
-i \frac{\partial H}{\partial \lambda} U(t)
-i H(t) \frac{\partial U}{\partial \lambda}.
\end{equation*}
\noindent We can multiply both sides from the left by $U^\dagger(t)$, which results in the following if we remember that the time-evolution is unitary:
\begin{equation*}
U^\dagger (t)\frac{\partial}{\partial \lambda} \frac{\mathrm{d}}{\mathrm{d}t} U(t) =
-i U^\dagger(t) \frac{\partial H}{\partial \lambda} U(t)
-i U^\dagger(t) H(t) \frac{\partial U}{\partial \lambda}.
\end{equation*}

Now, let us take the complex conjugate of the Schr\"{o}dinger equation and multiply it by $\partial U (t) /\partial \lambda$ from the right:
\begin{equation*}
\left[ \frac{\mathrm{d}}{\mathrm{d}t} U^\dagger (t) \right] \frac{\partial U}{\partial \lambda} = i U^\dagger (t) H(t) \frac{\partial U}{\partial \lambda}.
\end{equation*}
\noindent Adding both equations, we find:
\begin{equation*}
\frac{\mathrm{d}}{\mathrm{d} t} \left[ U^\dagger (t)\frac{\partial U}{\partial \lambda} \right] =
-i U^\dagger(t) \frac{\partial H}{\partial \lambda} U(t),
\end{equation*}
\noindent which, integrating both sides with respect to $t$, becomes:
\begin{equation*}
\left[ U^\dagger (t)\frac{\partial U }{\partial \lambda} - U^\dagger (0)\frac{\partial U}{\partial \lambda} \right] =
-i \int_0^t \mathrm{d}t^\prime U^\dagger(t^\prime) \frac{\partial H(t^\prime)}{\partial \lambda} U(t^\prime).
\end{equation*}

But, as $U(0)=1$, its partial derivative on the left-hand side is zero, leaving only:
\begin{equation*}
\frac{\partial U }{\partial \lambda} =
-i \int_0^t U(t) \mathrm{d}t^\prime U^\dagger(t^\prime) \frac{\partial H(t^\prime)}{\partial \lambda} U(t^\prime).
\end{equation*}
\noindent Furthermore, if the Hamiltonian has the form of Eq. (\ref{evolution}), we know that
\begin{equation*}
\frac{\partial H(t^\prime)}{\partial \lambda} = H_{\mathrm{int}}.
\end{equation*}
\section*{Appendix B: Time-independence of $\mathcal{E}_0 (t)$}

In this appendix, we are interested in verifying how a typical $\mathcal{E}_0 (t)$ varies in time, while making as few assumptions about the environment and the kind of error it generates as possible.

In general, the interaction-picture Hamiltonian of an object-system subject only to environmental noise can be written as (compare with Eq. (\ref{formamiltoniana})):

\begin{equation*}
H_I(t) = \sum_k S_k a_k e^{-i\omega_k t} + \text{h.c.},
\end{equation*}

\noindent where the operators $\{ S_k \}$ are time-independent and act only on the object-system.

As the trace terms only survive when they contain an equal number of $a_k$ and a $a_k^\dagger$, the trace of the error superoperator $\mathcal{E}_0 (t)$ when applied to the an initial-state density matrix $\rho$ can be written as:

\begin{eqnarray*}
\lefteqn{ \mathrm{Tr}_E \left\{ \mathcal{E}_0 (t) \rho \right\} = \;\;\;\;\;\;\;\;\;\;\;\;\; } \\
&&  \sum_k \int_0^1 du  \int_0^{1} dv \; e^{-i\omega_k (u-v)t}  \mathrm{Tr}_E \left\{ \left[  S_k a_k,  \left[ \rho,  S_k^\dagger a_k^\dagger \right] \right] \right\}  \\
&& + \sum_k \int_0^1 du  \int_0^{1} dv \; e^{i\omega_k (u-v)t} \mathrm{Tr}_E \left\{ \left[  S_k^\dagger a_k^\dagger,  \left[ \rho,  S_k a_k \right] \right] \right\}  \\
&& - \sum_k \int_0^1 du \int_0^u dv \; e^{-i\omega_k (u-v)t} \mathrm{Tr}_E \left\{ \left[ \left[ S_k a_k ,  S_k^\dagger a_k^\dagger \right] ,  \rho \right] \right\} \\
&& - \sum_k \int_0^1 du \int_0^u dv \; e^{i\omega_k (u-v)t} \mathrm{Tr}_E \left\{ \left[ \left[ S_k^\dagger a_k^\dagger ,  S_k a_k \right] ,  \rho \right] \right\}.
\end{eqnarray*}

The integrals in $u$ and $v$, which are all that matter for time dependence, result in:

\begin{eqnarray*}
\int_0^1 du  \int_0^{1} dv \; e^{-i\omega_k (u-v)t} &=& \frac{e^{-i \omega_k t} -1}{-i\omega_k t} \frac{e^{i \omega_k t}-1}{i\omega_k t} \\
&=& 2\frac{1-\cos(\omega_k t)}{(\omega_k t)^2} \\
&=& 1 + O(t^2)
\end{eqnarray*}

\noindent and

\begin{eqnarray*}
\int_0^1 du  \int_0^{u} dv \; e^{-i\omega_k (u-v)t} &=& \int_0^1 du \frac{1-e^{-i\omega_k u t}}{i\omega_k t} \\
&=& \frac{1}{i\omega_k t} - \frac{e^{-i\omega_k t}-1}{(\omega_k t)^2} \\
&=& \frac{1-i\omega_k t-e^{-i\omega_k t}}{(\omega_k t)^2} \\
&=& \frac{1}{2} + O(t).
\end{eqnarray*}

Therefore, it can be seen that there is no inverse-dependence with time. With time-independent terms dominating, we can be sure that the term that corresponds to the error in Eq. (\ref{final_res}) is not being underestimated. When multiplied by $\lambda^2 \Delta t^2$,  $\mathrm{Tr}_E \left\{ \mathcal{E}_0 (\Delta t) \rho \right\}$ will continue to be proportional to $\lambda^2 \Delta t^2$, besides introducing even smaller terms proportional to $\lambda^2 \Delta t^n$, with $n>2$.

\end{document}